# ELT Contributions to Tidal Disruption Events[1]

A Whitepaper Submitted to the Astro 2020 Decadal Survey Committee

J. Craig Wheeler (The University of Texas at Austin)
Rafaella Margutti (Northwestern University)
Ryan Chornock (Ohio University)
Dan Milisavljevic (Purdue University)
Maryam Modjaz (New York University)
Sung-Chul Yoon (Seoul National University)

Contact information for Primary Author:
J. Craig Wheeler
Department of Astronomy, The University of Texas at Austin
2515 Speedway, C1400
Austin, TX 78712-1205
512-471-6407
wheel@astro.as.utexas.edu

---

[1][1] Adapted from a chapter in the 2018 edition of the Science Book of the Giant Magellan Telescope Project.

**An ELT system with its large aperture and sensitive optical and near infrared imager spectrographs will make major contributions to the study of stars ripped apart by supermassive black holes.**

Stars that pass too close to supermassive black-holes (SMBHs) at the cores of their host galaxies can encounter death and be completely ripped apart by the SMBH tidal forces (Figure 1). The result of the tidal encounter is a flare of luminous radiation across the electromagnetic spectrum (X-ray to radio) powered by partial accretion of the stellar material onto the SMBH. Theoretically predicted in the seventies (e.g. Hills 1975, Frank & Rees 1976), these flares of radiation are now routinely observed and are known as Tidal Disruption Events (TDEs).

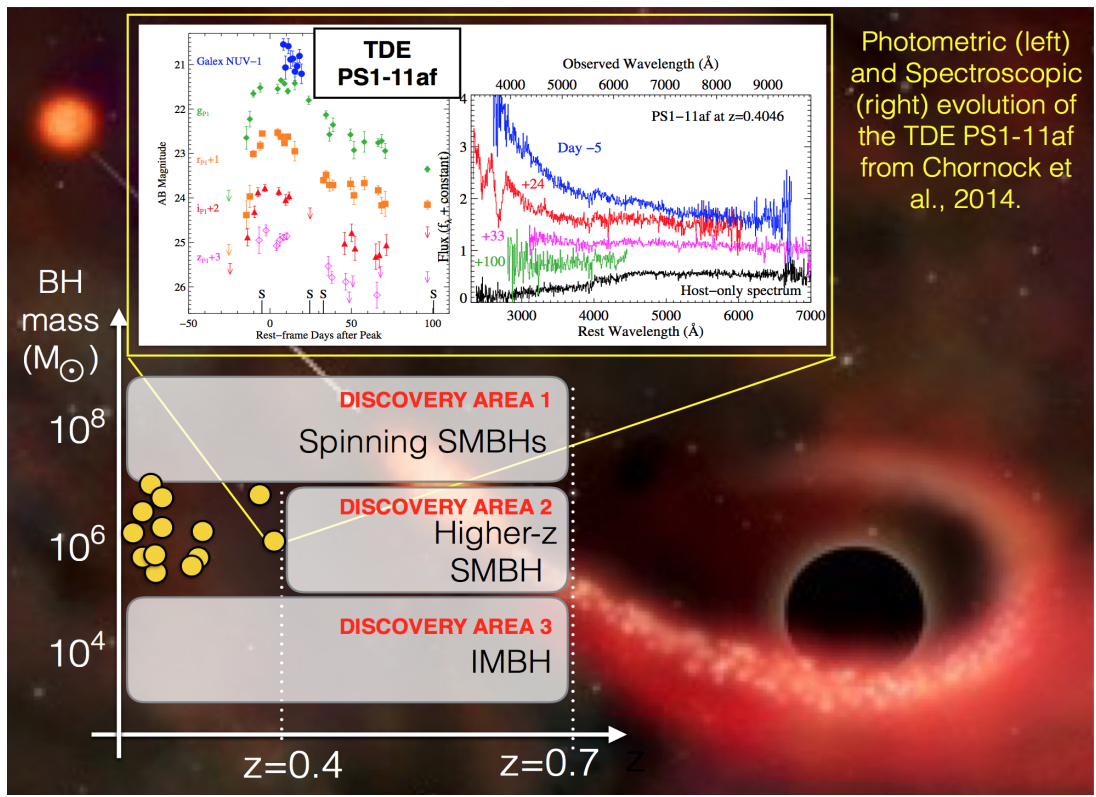

*Figure 1. Thanks to the spectroscopic sensitivity, spatial resolution, and spectral coverage it enables, ELT system observations will be able to reach a range of distances and masses that are inaccessible today in the study of supermassive black holes (SMBHs). Tidal disruption events (TDE) are luminous transients that originate from the disruption of stars by supermassive black holes (and perhaps intermediate mass black holes) in the cores of their host galaxies. This figure presents the current state of photometric and spectroscopic follow up (Chornock et al. 2014). An ELT system will greatly expand the fidelity of such data and the cosmic volume over which such systems can be studied.*

TDEs are interesting for many reasons. First, TDEs can be used as a marker for SMBHs that otherwise lie dormant and undetected in the centers of distant galaxies (>100 Mpc away), where they are too far away for the orbits of gas and stars around them to be resolved. Furthermore, TDEs are excellent probes of relativistic effects in regimes of strong gravity and provide a new means to measure SMBH masses *and* spins. Finally, TDEs are signposts of intermediate-mass BHs, binary BHs and recoiling BHs.

The first TDE candidates were detected in the nineties at X-ray wavelengths (by *ROSAT*), and later in UV (mainly by *GALEX*). TDEs are now routinely detected by current optical transient surveys (e.g. *PanSTARRS*, *ASAS-SN*, *PTF*; see Komossa 2015 for a review). Only a few TDEs have an associated radio counterpart so far. At the time of writing, only ~70 TDE candidates have been identified (https://tde.space). Even this small sample of events has greatly expanded our understanding of these transients.

- TDEs show a diverse phenomenology, with at least two broad classes identified so far. The first class includes high-energy TDEs associated with collimated relativistic outflows. A second class includes optical/UV TDEs with no signature of collimated outflows. At the time of writing, it is not clear what the key physical properties are that enable some SMBHs to launch relativistic jets at the time of stellar disruption.
- TDEs seem to show a preference for post-starburst galaxies. The reason for the overabundance of TDEs in post-starburst galaxies is unclear.
- The theoretically-predicted light-curve decay $\sim t^{(-5/3)}$ is only observed in a few TDEs. All TDEs with good data quality show deviations from this prediction.
- The spectral signatures of TDEs are also a puzzle. Some TDEs display He II emission lines, with no sign of H. Other TDEs instead do display H emission. The spectral properties of TDEs are the focus of intense theoretical modeling.

The limited understanding of the optical properties of TDEs, and the lack of a clear picture of how these map into the SMBH properties is significant, as it impairs our capabilities to realize the full potential of TDEs as probes of extreme gravity.

The limitations of current investigations can be traced to two key factors: statistics, and the limited range of redshifts probed (i.e. limited volume). TDEs are intrinsically rare (~$10^{-4}$-$10^{-5}$ events per galaxy per year). Like Type Ia supernovae, TDEs typically reach mag ~ -19 at maximum light. Different from supernovae, however, TDEs show a much longer-lived optical/UV emission, which can last for months. The much longer time scales of evolution make TDEs prime targets for LSST. Indeed, the LSST contribution to TDE studies will be substantial, with ~4000 TDEs discovered per year. More than 1000 TDE will be discovered per year above mag $24^{th}$, well within the reach of ELT optical/NIR spectroscopic capabilities.

An ELT system spectroscopic follow-up of LSST-discovered TDEs will enable the following *new* science:

(1) accurate census of dormant SMBH in the local Universe, with the potential to identify correlations of SMBH properties with host-galaxy properties (mass, bulge, type);
(2) demographics of dormant SMBH out to z~0.7, a redshift range currently not accessible;
(3) identification of intermediate-mass BHs in dwarf galaxies; and
(4) through spectropolarimetry, an ELT system will reveal the geometry of the emitting material.